\documentclass[12pt]{article}
\usepackage{epsfig}
\usepackage{amssymb}
\usepackage{amsmath}
\usepackage{amsfonts}

\newcommand{\R}{\mathbb{R}}
\newcommand{\C}{\mathbb{C}}

\newcommand{\be}{\begin{equation}}
\newcommand{\ee}{\end{equation}}
\newcommand{\bea}{\begin{eqnarray}}
\newcommand{\eea}{\end{eqnarray}}
\newcommand{\nn}{\nonumber}
\newcommand{\kt}{\rangle}
\newcommand{\br}{\langle}

\newcommand{\ed}{\end{document}}

\newcommand{\rx}{{\rm x}}
\newcommand{\ry}{{\rm y}}
\newcommand{\rp}{{\rm p}}

\newcommand{\rH}{{\rm H}}
\newcommand{\rX}{{\rm X}}
\newcommand{\rP}{{\rm P}}
\newcommand{\rh}{{\rm h}}
\newcommand{\rE}{{\rm E}}
\newcommand{\rv}{{\rm v}}

\newcommand{\np}{\newpage}

\begin{document}

\title{Application of Pseudo-Hermitian Quantum Mechanics to a ${\cal PT}$-Symmetric
Hamiltonian with a Continuum\\ of Scattering States}

\author{\\
Ali Mostafazadeh
\\
\\
Department of Mathematics, Ko\c{c} University,\\
34450 Sariyer, Istanbul, Turkey\\ amostafazadeh@ku.edu.tr}
\date{ }
\maketitle

\begin{abstract}
We extend the application of the techniques developed within the
framework of the pseudo-Hermitian quantum mechanics to study a
unitary quantum system described by an imaginary ${\cal
PT}$-symmetric potential $v(x)$ having a continuous real spectrum.
For this potential that has recently been used, in the context of
optical potentials, for modelling the propagation of
electromagnetic waves travelling in a wave guide half and half
filed with gain and absorbing media, we give a perturbative
construction of the physical Hilbert space, observables, localized
states, and the equivalent Hermitian Hamiltonian. Ignoring terms
of order three or higher in the non-Hermiticity parameter $\zeta$,
we show that the equivalent Hermitian Hamiltonian has the form
$\frac{p^2}{2m}+\frac{\zeta^2}{2}\sum_{n=0}^\infty\{\alpha_n(x),p^{2n}\}$
with $\alpha_n(x)$ vanishing outside an interval that is three
times larger than the support of $v(x)$, i.e., in 2/3 of the
physical interaction region the potential $v(x)$ vanishes
identically. We provide a physical interpretation for this unusual
behavior and comment on the classical limit of the system.
\end{abstract}
\begin{center}
~~~~~PACS numbers: 03.65.-w
\end{center}
\vspace{2mm}




\np

\section{Introduction}

During the past seven years there have appeared over two hundred
research papers on ${\cal PT}$-symmetric quantum systems. This was
initially triggered by the surprising observation of Bessis and
Zinn-Justin and its subsequent numerical verification by Bender
and his co-workers \cite{bender-PRL98-JMP99} that certain
non-Hermitian but ${\cal PT}$-symmetric Hamiltonians, such as
    \be
    H=p^2+x^2+i\epsilon x^3~~~~{\rm with}~~~~~~\epsilon\in\R^+,
    \label{cubic}
    \ee
have a purely real spectrum. This observation suggested the
possibility to use these Hamiltonians in the description of
certain quantum systems. Since the ${\cal PT}$-symmetry of a
non-Hermitian Hamiltonian $H$, i.e., the condition $[H,{\cal
PT}]=0$, did not ensure the reality of its spectrum, a crucial
task was to seek for the necessary and sufficient conditions for
the reality of the spectrum of a given non-Hermitian Hamiltonian
$H$. This was achieved in \cite{p2p3} where it was shown, under
the assumptions of the diagonalizability of $H$ and discreteness
of its spectrum, that the reality of the spectrum was equivalent
to the existence of a positive-definite inner product
$\br\cdot,\cdot\kt_+$ that rendered the Hamiltonian self-adjoint,
i.e., for any pair ($\psi,\phi$) of state vectors
$\br\psi,H\phi\kt_+=\br H\psi,\phi\kt_+$.

Another condition that is equivalent to the reality of the
spectrum of $H$ is that it can be mapped to a Hermitian
Hamiltonian $h$ via a similarity transformation
\cite{p2p3,jpa-2003}; there is an invertible Hermitian operator
$\rho$ such that
    \be
    H=\rho^{-1} h\, \rho.
    \label{similar}
    \ee
The positive-definite inner product $\br\cdot,\cdot\kt_+$ and the
operator $\rho$ entering (\ref{similar}) are determined by a
positive-definite operator $\eta_+$ according to
\cite{p2p3,jpa-2003}
    \bea
    &&\br\cdot,\cdot\kt_+:=\br\cdot|\eta_+\cdot\kt,
    \label{inn=inn}\\
    &&\rho=\sqrt{\eta_+},
    \label{rho}
    \eea
and the Hamiltonian satisfies the $\eta_+$-pseudo-Hermiticity
condition \cite{p1}:
    \be
    H^\dagger=\eta_+H\eta_+^{-1}.
    \label{ph}
    \ee
Here $\br\cdot|\cdot\kt$ stands for the standard $(L^2)$ inner
product that determines the (reference) Hilbert space ${\cal H}$
as well as the adjoint $H^\dagger$ of $H$,
\cite{jpa-2004c}.\footnote{The adjoint $A^\dagger$ of an operator
$A$ is the unique operator satisfying, for all $\psi,\phi\in{\cal
H}$, $\br\psi|A^\dagger\phi\kt=\br A\psi|\phi\kt$. $A$ is called
Hermitian if $A^\dagger=A$.}

It is this, so-called metric operator, $\eta_+$ that determines
the kinematic structure (the physical Hilbert space and the
observables) of the desired quantum system. Note however that
$\eta_+$ is not unique \cite{p4,jmp-2003,geyer-cjp}.\footnote{it
is only unique up to symmetries of the Hamiltonian,
\cite{jmp-2003}.} In \cite{p2p3} we have not only established the
existence of a positive definite metric operator $\eta_+$ and the
corresponding positive-definite inner product
$\br\cdot,\cdot\kt_+$ for a diagonalizable
Hamiltonian\footnote{For a treatment of non-diagonalizable
pseudo-Hermitian Hamiltonians see \cite{jmp-02d,ss,jmp-04}. Note
that diagonalizability of the Hamiltonian is a necessary condition
for applicability of the standard quantum measurement theory
\cite{jpa-2004c}. It is also necessary for the unitarity of the
time-evolution, for a non-diagonalizable Hamiltonian is never
Hermitian (its evolution operator is never unitary \cite{jmp-04})
with respect to a positive-definite inner product,
\cite{jmp-02d,ss}.} with a discrete real spectrum, but we have
also explained the role of antilinear symmetries such as ${\cal
PT}$ and offered a method for computing the most general $\eta_+$.
An alternative approach that yields a positive-definite inner
product for a class of ${\cal PT}$-symmetric models is that of
\cite{bender-PRL-2002}. As shown in \cite{jmp-2003,jpa-2005}, the
${\cal CPT}$-inner product proposed in \cite{bender-PRL-2002} is
identical with the inner product
$\br\cdot,\cdot\kt_+=\br\cdot|\eta_+\cdot\kt$ for a particular
choice of $\eta_+$.

Under the above mentioned conditions every Hamiltonian having a
real spectrum determines a set ${\cal U}_{H+}$ of
positive-definite metric operators. To formulate a consistent
unitary quantum theory having $H$ as its Hamiltonian, one needs to
choose an element $\eta_+$ of ${\cal
U}_{H+}$.\footnote{Alternatively one may choose sufficiently many
operators with real spectrum to construct a so-called irreducible
set of observables which subsequently fixes a metric operator
$\eta_+$, \cite{geyer}.} Each choice fixes a positive-definite
inner product $\br\cdot,\cdot\kt_+$ and defines the physical
Hilbert space ${\cal H}_{\rm phys}$ and the observables. The
latter are by definition \cite{critique} the operators $O$ that
are self-adjoint with respect to $\br\cdot,\cdot\kt_+$,
alternatively they are $\eta_+$-pseudo-Hermitian. These can be
constructed from Hermitian operators $o$ acting in ${\cal H}$
according to \cite{jpa-2004c}
    \be
    O=\rho^{-1} o\, \rho.
    \label{observable}
    \ee
In particular, one can define $\eta_+$-pseudo-Hermitian position
$X$ and momentum $P$ operators \cite{critique,jpa-2004c}, express
$H$ as a function of $X$ and $P$, and determine the underlying
classical Hamiltonian for the system by letting $\hbar\to 0$ in
the latter expression, \cite{jpa-2004c,p64}. Alternatively, one
may calculate the equivalent Hermitian Hamiltonian $h$ and obtain
its classical limit (again by letting $\hbar\to 0$).

Another application of the $\eta_+$-pseudo-Hermitian position
operator $X$ is in the construction of the physical localized
states:
    \be
    |\xi^{(x)}\kt:=\rho^{-1}|x\kt.
    \label{localized}
    \ee
These in turn define the physical position wave function,
$\Psi(x):=\br\xi^{(x)},\psi\kt_+=\br x|\rho|\psi\kt$, and the
invariant probability density,
    \be
    \varrho(x):=\frac{|\Psi(x)|^2}{\int_{-\infty}^\infty |
    \Psi(x)|^2dx}=\frac{|\br x|\rho|\psi\kt|^2}{
    \br\psi,\psi\kt_+},
    \label{density}
    \ee
for a given state vector $|\psi\kt$, \cite{jpa-2004c,p64}.

The above prescription for treating ${\cal PT}$-symmetric and more
generally pseudo-Hermitian Hamiltonians with a real spectrum has
been successfully applied in the study of the ${\cal
PT}$-symmetric square well in \cite{jpa-2004c} and the cubic
anharmonic oscillator (\ref{cubic}) in \cite{p64}.\footnote{See
also \cite{banerjee}.} Both these systems have a discrete
nondegenerate energy spectrum, and the results of \cite{p1,p2p3}
are known to apply to them. The aim of the present paper is to
seek whether these results (in particular the construction method
for $\eta_+$) may be used for treating a system with a continuous
spectrum.\footnote{The question whether the theory of
pseudo-Hermitian operators as outlined in \cite{p1,p2p3} is
capable of treating a system having scattering states was posed to
the author by Zafar Ahmed during the 2nd International Workshop on
Pseudo-Hermitian Hamiltonians in Quantum Physics, held in Prague,
June 14-16, 2004.} This question is motivated by the desire to
understand field-theoretical analogues of ${\cal PT}$-symmetric
systems which should admit an $S$-matrix formulation. Furthermore,
there are some basic questions related to the nonlocal nature of
the Hermitian Hamiltonian $h$ and the pseudo-Hermitian observables
such as $X$ and $P$ especially for ${\cal PT}$-symmetric
potentials with a compact support (i.e., potentials vanishing
outside a compact region).

To achieve this aim we will focus our attention on a simple toy
model recently considered as an effective model arising in the
treatment of the electromagnetic waves travelling in a planar slab
waveguide that is half and half filed with gain and absorbing
media, \cite{ruschhaupt}. This model has a standard Hamiltonian,
    \be
    H=\frac{p^2}{2m}+v(x),
    \label{H}
    \ee
and a ${\cal PT}$-symmetric imaginary potential,
    \be
    v(x):=i\zeta[\theta(x+\mbox{\small$\frac{L}{2}$})+
    \theta(x-\mbox{\small$\frac{L}{2}$})-2\,\theta(x)]=
    \left\{\begin{array}{ccc}
    0&{\rm for}&|x|\geq \mbox{\small$\frac{L}{2}$}~~{\rm or}~~x=0\\
    i\zeta &{\rm for}& x\in (-\mbox{\small$\frac{L}{2}$},0)\\
    -i\zeta&{\rm for}& x\in (0,\mbox{\small$\frac{L}{2}$}),
    \end{array}\right.
    \label{v}
    \ee
where $L\in (0,\infty)$ is a length scale, $\zeta\in [0,\infty)$
determines the degree of non-Hermiticity of the system, and
$\theta$ is the step function:
    \be
    \theta(x):=\left\{\begin{array}{ccc}
    0&{\rm for}&x<0\\
    \mbox{\small$\frac{1}{2}$}&{\rm for}&x=0\\
    1&{\rm for}& x>0.
    \end{array}\right.
    \label{theta}
    \ee
The Hamiltonian~(\ref{H}) differs from a free particle Hamiltonian
only within
$(-\mbox{\small$\frac{L}{2}$},\mbox{\small$\frac{L}{2}$})$ where
it coincides with the Hamiltonian for the ${\cal PT}$-symmetric
square well \cite{sw-bmq,jpa-2004c}.

It is important to note that unlike in \cite{ruschhaupt} we will
consider the potential (\ref{v}) as defining a fundamental
(non-effective) quantum system having a unitary time-evolution
(and $S$-matrix). Therefore our approach will be completely
different from that pursued in \cite{ruschhaupt} and the earlier
studies of effective (optical) non-Hermitian Hamiltonians,
\cite{other}.

To the best of author's knowledge, the only other non-Hermitian
Hamiltonian with a continuous (and doubly degenerate) spectrum
that is shown to admit a similar treatment is the one arising in
the two-component formulation of the free Klein-Gordon equation
\cite{cqg,kg}. Compared to (\ref{H}), this Hamiltonian defines a
technically much simpler system to handle, because it is
essentially a tensor product of an ordinary Hermitian Hamiltonian
and a $2\times 2$ matrix pseudo-Hermitian Hamiltonian.

\section{Metric Operator}

The essential ingredient of our approach is the metric operator
$\eta_+$. For a diagonalizable Hamiltonian with a discrete
spectrum it can be expressed as
    \be
    \eta_+=\sum_n \sum_{a=1}^{d_a} |\phi_n,a\kt\br\phi_n,a|,
    \label{eta}
    \ee
where $n$, $a$, and $d_n$ are a spectral label, a degeneracy
label, and the multiplicity (degree of degeneracy) for the
eigenvalue $E_n$ of $H$, respectively, and $\{|\phi_n,a\kt\}$ is a
complete set of eigenvectors of $H^\dagger$ that together with the
eigenvectors $|\psi_n,a\kt$ of $H$ form a biorthonormal system,
\cite{p1,p2p3}.

Now, consider a diagonalizable Hamiltonian with a purely
continuous doubly degenerate real spectrum $\{E_k\}$, where $k\in
(0,\infty)$. We will extend the application of (\ref{eta}) to this
Hamiltonian by changing $\sum_n\cdots $ to $\int dk\cdots$. This
yields
    \be
    \eta_+=\int_0^\infty dk\,(|\phi_k,+\kt\br\phi_k,+|+
    |\phi_k,-\kt\br\phi_k,-|),
    \label{eta=}
    \ee
where we have used $\pm$ as the values of the degeneracy label
$a$, \cite{kg}. The biorthonormal system
$\{|\psi_k,a\kt,|\phi_k,a\kt\}$ satisfies
    \bea
    H|\psi_k,a\kt&=&E_k|\psi_k,a\kt,~~~~~~~
    H^\dagger|\phi_k,a\kt=E_k|\phi_k,a\kt,
    \label{eg-va}\\
    \br\phi_k,a|\psi_\ell,b\kt&=&\delta_{ab}\delta(k-\ell),~~~~~~~
    \int_0^\infty (|\psi_k,+\kt\br\phi_k,+|+
    |\psi_k,-\kt\br\phi_k,-|)\,dk=1,
    \label{complete}
    \eea
where $\delta_{ab}$ and $\delta(k)$ stand for the Kronecker and
Dirac delta functions, respectively, $k\in(0,\infty)$, and
$a,b\in\{-,+\}$,

We define the eigenvalue problem for the Hamiltonian (\ref{H})
using the oscillating (plane wave) boundary conditions at
$x=\pm\infty$ similarly to the free particle case which
corresponds to $\zeta=0$. To simplify the calculation of the
eigenvectors we first introduce the following dimensionless
quantities.
    \bea
    &&\rx:=(\mbox{\small$\frac{2}{L}$})\,x,~~~~~~
    \rp:=(\mbox{\small$\frac{L}{2\hbar}$})\,p,~~~~~~
    Z:=(\mbox{\small$\frac{mL^2}{2\hbar^2}$})\,\zeta,~~~~~~
    \rH:=(\mbox{\small$\frac{mL^2}{2\hbar^2}$})\,H=\rp^2+\rv(\rx),
    \label{scale1}\\
    &&\rv(\rx):=i Z[\theta(\rx+1)+
    \theta(\rx-1)-2\,\theta(\rx)]=
    \left\{\begin{array}{ccc}
    0&{\rm for}&|\rx|\geq 1~~{\rm or}~~\rx=0\\
    i Z &{\rm for}& \rx\in (-1,0)\\
    -i Z&{\rm for}& \rx\in (0,1).
    \end{array}\right.
    \label{r-v}
    \eea

The eigenvalue problem for the scaled Hamiltonian $\rH$
corresponds to the solution of the differential equation
    \be
    \left[-\frac{d^2}{d\rx^2}+\rv(\rx)-\rE_k\right]\psi(\rx)=0,
    \label{DE}
    \ee
that is subject to the condition that $\psi$ is a differentiable
function at the discontinuities $\rx=-1,0,1$ of $\rv$. Introducing
$\psi_1:(-\infty,-1]\to\C$, $\psi_-:[-1,0]\to\C$,
$\psi_+:[0,1]\to\C$, and $\psi_2:[1,\infty)\to\C$ according to
    \be
    \psi(\rx)=:\left\{\begin{array}{ccc}
    \psi_1(\rx)&{\rm for}&\rx\in (-\infty,-1]\\
    \psi_-(\rx)&{\rm for}&\rx\in [-1,0]\\
    \psi_+(\rx)&{\rm for}&\rx\in [0,1]\\
    \psi_2(\rx)&{\rm for}&\rx\in [1,\infty),
    \end{array}\right.
    \label{psi}
    \ee
we have
    \bea
    \psi_1(-1)&=&\psi_-(-1),~~~~~~~\psi'_1(-1)=\psi'_-(-1),
    \label{c1}\\
    \psi_-(0)&=&\psi_+(0),~~~~~~~\psi'_-(0)=\psi'_+(0),
    \label{c2}\\
    \psi_+(1)&=&\psi_2(1),~~~~~~~\psi'_+(1)=\psi'_2(1).
    \label{c3}
    \eea
Now, imposing the plane-wave boundary condition at $\rx=\pm\infty$
and demanding that the eigenfunctions $\psi$ be ${\cal
PT}$-invariant, which implies
    \be
    \psi_-(0)=\psi_+(0)^*,~~~~~~~~~~~~
    \psi_-'(0)=-\psi_+'(0),
    \label{c4}
    \ee
we find $E_k=k^2$, i.e., the spectrum is real positive and
continuous, and
    \be
    \psi_1(\rx)=A_1 e^{ik\rx}+B_1e^{-ikx},~~~~
    \psi_2(\rx)=A_2 e^{ik\rx}+B_2e^{-ikx},~~~~
    \psi_\pm(\rx)=A_\pm e^{ik_\pm\rx}+B_\pm e^{-ik_\pm\rx},
    \label{psi=}
    \ee
where
    \bea
    &&k_\pm:=\sqrt{k^2\pm iZ},
    \label{k-pm}\\
    &&A_1=A_3^*=\frac{e^{ik}}{\sqrt{2\pi}}\left[
    L_-(k)u+K_-(k)v\right],~~~~~
    B_1=B_3^*=\frac{e^{-ik}}{\sqrt{2\pi}}\left[
    L_-(-k)u+K_-(-k)v\right],~~~~~~~~
    \label{A1,B1}\\
    &&L_-(k):=\frac{1}{2}\left(\cos k_--\frac{ik_-\sin
    k_-}{k}\right),~~~~~
    K_-(k):=\frac{1}{2}\sqrt{\frac{k_+}{k_-}}
    \left(\frac{k_-\cos k_-}{k}-i\sin
    k_-\right),
    \label{L-K}\\
    &&A_\pm=\frac{1}{\sqrt {8\pi}}
    \left[u+\left(\frac{k_+}{k_-}\right)^{\pm 1/2}
    v\right],~~~~~~~~~~
    B_\pm=\frac{1}{\sqrt {8\pi}}
    \left[u-\left(\frac{k_+}{k_-}\right)^{\pm 1/2}
    v\right],
    \label{AB-pm}
    \eea
and $u,v\in\R$ are arbitrary constants (possibly depending on $k$
and/or $Z$ and not both vanishing).

The presence of the free parameters $u$ and $v$ is an indication
of a double degeneracy of the eigenvalues $\rE_k=k^2$. We will
select $u$ and $v$ in such a way as to ensure that in the limit
$Z\to 0$ we recover the plane-wave solutions of the free particle
Hamiltonian, i.e., we demand $\lim_{Z\to 0}\psi(\rx) = e^{\pm
ik\rx}/\sqrt{2\pi}$. This condition is satisfied if we set
    \be
    u=1,~~~~~~~~~v=\pm 1.
    \label{u-v}
    \ee
In the following we use the superscript $\pm$ to identify the
value of a quantity obtained by setting $u=1$ and $v=\pm 1$. In
this way we introduce $A_1^\pm,B_1^\pm,A_2^\pm,B_2^\pm, A_\pm^\pm,
B_\pm^\pm$, and $\psi^\pm$. The latter define the basis
(generalized \cite{bohm-qm}) eigenvectors $|\psi_k,\pm\kt$ by
$\br\rx|\psi_k,\pm\kt:=\psi^\pm(\rx)$.

The next step is to obtain $|\phi_k,\pm\kt$. In view of the
identity $\rH^\dagger=\rH|_{Z\to -Z}$, we can easily obtain the
expression for the eigenfunctions $\phi$ of $H^\dagger$.
Introducing
    \be
    \phi(\rx)=:\left\{\begin{array}{ccc}
    \phi_1(\rx)&{\rm for}&\rx\in (-\infty,-1]\\
    \phi_-(\rx)&{\rm for}&\rx\in [-1,0]\\
    \phi_+(\rx)&{\rm for}&\rx\in [0,1]\\
    \phi_2(\rx)&{\rm for}&\rx\in [1,\infty),
    \end{array}\right.
    \label{phi}
    \ee
we have
    \be
    \phi_1(\rx)=C_1 e^{ik\rx}+D_1e^{-ikx},~~~~
    \phi_2(\rx)=C_2 e^{ik\rx}+D_2e^{-ikx},~~~~
    \phi_\pm(\rx)=C_\pm e^{ik_\mp\rx}+D_\pm e^{-ik_\mp\rx},
    \label{phi=}
    \ee
where
    \bea
    &&C_1=C_3^*=\frac{e^{ik}}{\sqrt{2\pi}}\left[
    L_+(k)r+K_+(k)s\right],
    ~~~~
    D_1=D_3^*=\frac{e^{-ik}}{\sqrt{2\pi}}\left[
    L_+(-k)r+K_+(-k)s\right],~~~~~~~~~~
    \label{D1}\\
    &&L_+(k):=L_-(-k)^*,~~~~~~~~~~~~~~~~~~~~~
    K_+(k):=-K_-(-k)^*,
    \label{LK-plus}\\
    &&C_\pm=\frac{1}{\sqrt {8\pi}}
    \left[r+\left(\frac{k_+}{k_-}\right)^{\mp 1/2}
    s\right],~~~~~~~~~~
    D_\pm=\frac{1}{\sqrt {8\pi}}
    \left[r-\left(\frac{k_+}{k_-}\right)^{\mp 1/2}
    s\right],
    \label{CD-pm}
    \eea
and $r,s\in\R$ are (possibly $k$- and/or $Z$-dependent) parameters
that are to be fixed by imposing the biorthonormality condition
(\ref{complete}). The latter is equivalent to a set of four
(complex) equations (corresponding to the four possible choice for
the pair of indices $(a,b)$ in the first equation in
(\ref{complete})) which are to be solved for the two real unknowns
$r$ and $s$. This together with the presence of the delta function
in two of these equation make the existence of a solution quite
nontrivial.

We checked these equations by expanding all the quantities in
powers of the non-Hermiticity parameter $Z$ up to (but not
including) terms of order two and found after a long and tedious
calculation (partly done using Mathematica) that indeed all four
of these equations are satisfied, if we set $r=u=1$ and $s=v=\pm
1$. Again we will refer to this choice using superscript $\pm$. In
particular, we have $\phi^\pm=\psi^\pm|_{Z\to -Z}$ and $
\br\rx|\phi_k,\pm\kt:=\phi^\pm(\rx)$.

Having obtained $|\phi_k,\pm\kt$ we are in a position to calculate
the metric operator (\ref{eta=}). We carried out this calculation
using first order perturbation theory in $Z$. It involved
expanding the $\phi_1^\pm(\rx), \psi_2^\pm(\rx)$, and
$\psi_\pm^\pm(\rx)$ in powers of $Z$, substituting the result in
    \be
    \br \rx|\eta_+|\ry\kt=\int_0^\infty
    [\phi^+(\ry)^*\,\phi^+(\rx)+\phi^-(\ry)^*\,\phi^-(\rx)]dk
    \label{eta-int}
    \ee
which follows from (\ref{eta=}), and using the identities:
    \be
    \int_{-\infty}^\infty e^{i a k}dk=2\pi\delta(a),
    ~~~\int_{-\infty}^\infty \frac{e^{i a k}}{k}\,dk=
    i\pi\,{\rm sign}(a),~~~
    \int_{-\infty}^\infty \frac{e^{i a k}-e^{i b k}}{k^2}\,dk=
    \pi(|b|-|a|)
    \label{identities}
    \ee
(where $a,b\in\R$ and ${\rm sign}(a):=\theta(a)-\theta(-a)$) to
perform the integral over $k$ for all 16 possibilities for the
range of values of the pair of independent variables $(\rx,\ry)$
in (\ref{eta-int}). To simplify the presentation of the result, we
introduce: $I_1:=(-\infty,-1)$, $I_-:=(-1,0)$, $I_+:=(0,1)$,
$I_2:=(1,\infty)$ and define the functions ${\rm
E}_{\mu,\nu}:I_\mu\times I_\nu\to\C$ by
    \[{\cal E}_{\mu,\nu}(\rx,\ry):=\br \rx|\eta_+|\ry\kt~~~~
    {\rm for~all}~~~~\rx\in I_\mu,~\ry\in
    I_\nu,~~~\mu,\nu\in\{1,-,+,2\}.\]
Then after a very long calculation we find
    \bea
    {\cal E}_{1,1}(\rx,\ry)&=&\delta(\rx-\ry)+
    {\mbox{\small$\frac{i}{2}$}}\,{\rm sign}(\rx-\ry)\,Z
    +{\cal O}(Z^2),
    \label{e11}\\
    {\cal E}_{-,1}(\rx,\ry)&=&
    {\mbox{\small$\frac{i}{8}$}}\,
    (2-\rx-\ry-|\rx+\ry+2|)\,Z+{\cal O}(Z^2),
    \label{e12}\\
    {\cal E}_{+,1}(\rx,\ry)&=&
    {\mbox{\small$\frac{i}{8}$}}\,
    (2-\rx-\ry-|\rx+\ry+2|)\,Z+{\cal O}(Z^2),
    \label{e13}\\
    {\cal E}_{2,1}(\rx,\ry)&=&{\mbox{\small$\frac{i}{8}$}}\,
    (4+2|\rx+\ry|-|\rx+\ry+2|-|\rx+\ry-2|)\,Z+{\cal O}(Z^2),
    \label{e14}\\
    {\cal E}_{1,-}(\rx,\ry)&=&-{\mbox{\small$\frac{i}{8}$}}\,
    (2-\rx-\ry-|\rx+\ry+2|)\,Z+{\cal O}(Z^2),
    \label{e21}\\
    {\cal E}_{-,-}(\rx,\ry)&=&\delta(\rx-\ry)-
    {\mbox{\small$\frac{i}{4}$}}\,{\rm sign}(\rx-\ry)
    (\rx+\ry)\,Z+{\cal O}(Z^2),
    \label{e22}\\
    {\cal E}_{+,-}(\rx,\ry)&=&{\mbox{\small$\frac{i}{4}$}}\,
    |\rx+\ry|\,Z+{\cal O}(Z^2),
    \label{e23}\\
    {\cal E}_{2,-}(\rx,\ry)&=&{\mbox{\small$\frac{i}{8}$}}\,
    (2+\rx+\ry-|\rx+\ry-2|)\,Z+{\cal O}(Z^2),
    \label{e24}\\
    {\cal E}_{1,+}(\rx,\ry)&=&-{\mbox{\small$\frac{i}{8}$}}\,
    (2-\rx-\ry-|\rx+\ry+2|)\,Z+{\cal O}(Z^2),
    \label{e31}\\
    {\cal E}_{-,+}(\rx,\ry)&=&-{\mbox{\small$\frac{i}{4}$}}\,
    |\rx+\ry|\,Z+{\cal O}(Z^2),
    \label{e32}\\
    {\cal E}_{+,+}(\rx,\ry)&=&\delta(\rx-\ry)+
    {\mbox{\small$\frac{i}{4}$}}\,{\rm sign}(\rx-\ry)
    (\rx+\ry)\,Z+{\cal O}(Z^2),
    \label{e33}\\
    {\cal E}_{2,+}(\rx,\ry)&=&{\mbox{\small$\frac{i}{8}$}}\,
    (2+\rx+\ry-|\rx+\ry-2|)\,Z+{\cal O}(Z^2),
    \label{e34}\\
    {\cal E}_{1,2}(\rx,\ry)&=&-{\mbox{\small$\frac{i}{8}$}}\,
    (4+2|\rx+\ry|-|\rx+\ry+2|-|\rx+\ry-2|)\,Z+{\cal O}(Z^2),
    \label{e41}\\
    {\cal E}_{-,2}(\rx,\ry)&=&-{\mbox{\small$\frac{i}{8}$}}\,
    (2+\rx+\ry-|\rx+\ry-2|)\,Z+{\cal O}(Z^2),
    \label{e42}\\
    {\cal E}_{+,2}(\rx,\ry)&=&-{\mbox{\small$\frac{i}{8}$}}\,
    (2+\rx+\ry-|\rx+\ry-2|)\,Z+{\cal O}(Z^2),
    \label{e43}\\
    {\cal E}_{2,2}(\rx,\ry)&=&\delta(\rx-\ry)+
    {\mbox{\small$\frac{i}{2}$}}\,{\rm sign}(\rx-\ry)\,Z
    +{\cal O}(Z^2),
    \label{e44}
    \eea
where ${\cal O}(Z^2)$ stands for terms of order two and higher in
powers of $Z$. It is quite remarkable that we can obtain from
(\ref{e11}) -- (\ref{e44}) a single formula for
$\br\rx|\eta_+|\ry\kt$ which is valid for all $\rx,\ry\in\R$,
namely
    \be
    \br\rx|\eta_+|\ry\kt=\delta(\rx-\ry)+
    {\mbox{\small$\frac{i}{8}$}}\,
    (4+2|\rx+\ry|-|\rx+\ry+2|-|\rx+\ry-2|)\,
    {\rm sign}(\rx-\ry)\,Z+{\cal O}(Z^2).
    \label{eta-y-x=}
    \ee
Note that $\br\rx|\eta_+|\ry\kt^*=\br\ry|\eta_+|\rx\kt$ which is
consistent with the Hermiticity of $\eta_+$.

\section{Physical Observables and Localized States}

The physical observables of the system described by the
Hamiltonian (\ref{H}) are obtained from the Hermitian operators
acting in ${\cal H}=L^2(\R)$ by the similarity transformation
(\ref{observable}). This equation involves the positive square
root $\rho$ of $\eta_+$ which takes the form \cite{p64}
    \be
    \rho^{\pm 1}=e^{\mp Q/2},
    \label{rho=}
    \ee
if we express $\eta$ in the exponential form
    \be
    \eta_+=e^{-Q}.
    \label{exp}
    \ee
In view of (\ref{rho=}) and the Backer-Campbell-Hausdorff
identity,
    \be
    e^{-A}B\,e^A=B+[B,A]+\frac{1}{2!}\,[[B,A],A]+\cdots
    \label{cbh}
    \ee
(where $A$ and $B$ are linear operators), physical observables
(\ref{observable}) satisfy \cite{p64}:
    \be
    O=o-\frac{1}{2}\,[o,Q]+\frac{1}{8}\,[[o,Q],Q]+\cdots.
    \label{O-calc}
    \ee

If we expand $\eta_+$ and $Q$ in powers of $Z$,
    \be
    \eta_+=1+\sum_{\ell=1}^\infty
    \eta_{+_\ell}Z^\ell,~~~~~~~
    Q=\sum_{\ell=1}^\infty
    Q_\ell Z^\ell,
    \label{expand}
    \ee
where $\eta_{+_\ell}$ and $Q_\ell$ are $Z$-independent Hermitian
operators, we find using (\ref{exp}) that
    \be
    Q_1=-\eta_{+_1},~~~~~~Q_2=-\eta_{+_2}+\frac{1}{2}\,
    \eta_{+_1}^2.
    \label{Q12}
    \ee
Combining this relation with (\ref{O-calc}), we have
    \be
    O=o-\frac{1}{2}[o,Q_1]\,Z+\frac{1}{8}(-4[o,Q_2]+
    [[o,Q_1],Q_1])\,Z^2+
    {\cal O}(Z^3).
    \label{pert-1}
    \ee

In the following we calculate the $\eta_+$-pseudo-Hermitian
position ($X$) and momentum $(P)$ operators, \cite{p64}, up to
(but no including) terms of order $Z^2$. This is because so far we
have only calculated $\eta_{+_1}$ which in view of
(\ref{eta-y-x=}) satisfies
    \be
    \br\rx|\eta_{+_1}|\ry\kt={\mbox{\small$\frac{i}{8}$}}\,
    (4+2|\rx+\ry|-|\rx+\ry+2|-|\rx+\ry-2|)\,
    {\rm sign}(\rx-\ry),~~~~~~~~\forall \rx,\ry\in\R.
    \label{eta-1-xy}
    \ee
Substituting the scaled position ($\rx$) and momentum $(\rp)$
operator for $o$ in (\ref{pert-1}), using (\ref{eta-1-xy}), and
doing the necessary algebra, we find
    \bea
    \br\rx|\rX|\ry\kt&=&\rx\,\delta(\rx-\ry)+
    {\mbox{\small$\frac{i}{16}$}}\,
    (4+2|\rx+\ry|-|\rx+\ry+2|-|\rx+\ry-2|)|\rx-\ry|\,Z+{\cal
    O}(Z^2),~~~~~
    \label{X=}\\
    \br\rx|\rP|\ry\kt&=&-i\partial_{\rx}\,\delta(\rx-\ry)+{\mbox{\small$\frac{1}{8}$}}\,
    [2\,{\rm sign}(\rx+\ry)-{\rm sign}(\rx+\ry+2)
    \nn\\
    &&\hspace{3.5cm}-
    {\rm sign}(\rx+\ry-2)]\:{\rm sign}(\rx-\ry)\,Z+{\cal
    O}(Z^2),
    \label{P=}
    \eea
where $\rX:=2 X/L$ and $\rP:=L P/(2\hbar)$ are dimensionless
$\eta_+$-pseudo-Hermitian position and momentum operators,
respectively.

As seen from (\ref{X=}), both $\rX$ and $\rP$ are manifestly
nonlocal and non-Hermitian (but pseudo-Hermitian) operators.
Furthermore,
    \[ \br\rx|\rP|\ry\kt=\br\rx|\rp|\ry\kt+{\cal O}(Z^2)~~~~
    {\rm for}~~~~\rx\notin [-1,1]~~{\rm and}~~\ry\notin [-1,1].\]

If we scale back the relevant quantities in (\ref{X=}) and
(\ref{P=}) according to (\ref{scale1}), we find
    \bea
    \br x|X|y\kt&=& x\,\delta(x-y)+
    {\mbox{\small$\frac{im}{4\hbar^2}$}}\,
    (2L+2|x+y|-|x+y+L|-|x+y-L|)|x-y|\,\zeta+{\cal
    O}(\zeta^2),~~~~~
    \label{X=scale}\\
    \br x|P|y\kt&=&-i\hbar\partial_x
    \delta(x-y)+{\mbox{\small$\frac{m}{4\hbar}$}}\,
    [2\,{\rm sign}(x+y)-{\rm sign}(x+y+L)\nn\\
    &&\hspace{4cm}-{\rm sign}(x+y-L)]\,
    {\rm sign}(x-y)\,\zeta+{\cal O}(\zeta^2).
    \label{P=scale}
    \eea
Again note that the contributions of order $\zeta$ to $P$ vanish,
if both $x$ and $y$ take values outside
$[-\mbox{\small$\frac{L}{2}$},\mbox{\small$\frac{L}{2}$}]$.

Next, we compute the localized states $\xi^{(x)}$ of the system.
The corresponding state vectors are defined by (\ref{localized}).
Using this equation as well as (\ref{rho=}), (\ref{expand}),
(\ref{Q12}), (\ref{eta-1-xy}), and (\ref{scale1}) we have the
following expression for the $x$-representation of a localized
state $\xi^{(y)}$ centered at $y\in\R$.
    \be
    \br x|\xi^{(y)}\kt=\delta(x-y)-
    \frac{im\zeta}{8\hbar^2}\,
    (2L+2|x+y|-|x+y+L|-|x+y-1|)\,
    {\rm sign}(x-y)+{\cal O}(\zeta^2).
    \label{localized-wf}
    \ee
Because the linear term in $\zeta$ is imaginary, the presence of a
weak non-Hermiticity only modifies the usual (Hermitian) localized
states by making them complex (non-real) while keeping their real
part intact. Note however that for a fixed $y$ the imaginary part
of $\br x|\xi^{(y)}\kt$ does not tend to zero as $|x-y|\to\infty$.
This observation which seems to be in conflict with the usual
notion of localizability has a simple explanation. Because the
usual $x$ operator is no longer an observable, it does not
describe the position of the particle. This is done by the
pseudo-Hermitian position operator $X$; it is the physical
position wave function $\Psi(x):=\br\xi^{(x)},\psi\kt_+$ that
defines the probability density of localization in space
(\ref{density}). The physical position wave function for the
localized state $\xi^{(y)}$ is given by
$\br\xi^{(x)},\xi^{(y)}\kt_+= \br x|y\kt=\delta(x-y)$ which is the
expected result.

In summary, the notion of localizability in space is directly
linked with the choice of the physical position operator. An
important by-product of the recent intensive investigation of
non-Hermitian ${\cal PT}$-symmetric systems is the realization of
the fact that one my formulate a unitary quantum system for which
the choice of the Hilbert space and observables, particularly the
position operator, is not a priori fixed.

\section{Equivalent Hermitian Hamiltonian and Classical
Limit}

The calculation of the equivalent Hermitian Hamiltonian $h$ for
the Hamiltonian (\ref{H}) is similar to that of the physical
observables. In view of (\ref{similar}), (\ref{rho=}),
(\ref{cbh}), (\ref{expand}), and the last equation in
(\ref{scale1}) which we express as
    \be
    \rH=\rp^2+i\nu(\rx)Z~~~~{\rm with}~~~~\nu(\rx):=
    \theta(\rx+1)+\theta(\rx-1)-2\theta(\rx),
    \label{scale-H}
    \ee
we have
    \bea
    \rh&=&\rp^2+h_1Z+h_2Z^2+{\cal O}(Z^3),
    \label{h-expand}\\
    \rh_1&:=&i\nu(\rx)+\frac{1}{2}[\rp^2,Q_1],
    \label{h1}\\
    \rh_2&:=&\frac{1}{8}\{
    4[\rp^2,Q_2]+4i[\nu(\rx),Q_1]+[[\rp^2,Q_1],Q_1]\}.
    \label{h2}
    \eea
where
    \be
    \rh:=\rho\,\rH\rho^{-1}=mL^2 h/(2\hbar^2)
    \label{rh}
    \ee
is the dimensionless Hermitian Hamiltonian associated with $\rH$.

Next, we substitute (\ref{Q12}) and (\ref{eta-1-xy}) in the
identity
    \[
    \br\rx|[\rp^2,Q_1]|\rv\kt=
    (\partial_{\ry}^2-\partial_{\rx}^2)\br\rx|Q_1|\ry\kt,\]
and perform the necessary algebra. We then find
$\br\rx|[\rp^2,Q_1]|\rv\kt=-2i\nu(\rx)\delta(\rx-\ry)$. Therefore,
    \be
    [\rp^2,Q_1]=-2i\nu(\rx),
    \label{id1}
    \ee
and in view of (\ref{h1})
    \be
    \rh_1=0.
    \label{h1=}
    \ee
This was actually to be expected, for both the operators appearing
on the right-hand side of (\ref{h1}) are anti-Hermitian, while its
left-hand side is Hermitian. The fact that an explicit calculation
of the right-hand side of (\ref{h1}) yields the desired result,
namely (\ref{h1=}), is an important check on the validity of our
calculation of $\eta_{+_1}$. It may also be viewed as an
indication of the consistency and general applicability of our
method, that was initially formulated for systems with a discrete
spectrum \cite{jpa-2004c,p64}.

According to (\ref{h1=}),
    \be
    \rh=\rp^2+\rh_2Z^2+{\cal O}(Z^3).
    \label{h=h2}
    \ee
Hence, in order to obtain a better understanding of the nature of
the system described by the Hamiltonian $H$, we need to calculate
$\rh_2$. Equations (\ref{h2}) and (\ref{Q12}) suggest that this
calculation demands a complete knowledge of $\eta_{+_2}$ which in
turn requires the calculation of $\br\rx|\eta_{+_2}|\ry\kt$ for
all 16 possibilities for the ranges of $\rx$ and $\ry$. This is an
extremely lengthy calculation in which one must deal with infinite
integrals of the form $\int_{-\infty}^\infty e^{iak}/k^n dk$ with
$n=2,3,4$.\footnote{These may be easily regularized as is well
known in typical field theory calculations.} We will not include
the result of this calculation here, not only because it is too
lengthy but most importantly because, as we will show in the
following, the knowledge of $\br\rx|\eta_{+_1}|\ry\kt$ turns out
to be sufficient for the calculation of $\rh_2$. To see this we
first employ (\ref{id1}) to express $\rh_2$ in the form
    \be
    \rh_2=\frac{1}{4}\,\left( 2[\rp^2,Q_2]+i[\nu(\rx),Q_1]\right).
    \label{id2}
    \ee
Now, we recall that $\rp^2$, $Q_2$, $\nu(\rx)$ and $Q_1$ are all
Hermitian operators. Therefore $[\rp^2,Q_2]$ and $i[\nu(\rx),Q_1]$
are respectively anti-Hermitian and Hermitian. In view of
(\ref{id2}) and the Hermiticity of $\rh_2$, this implies that
    \be
    [\rp^2,Q_2]=0.
    \label{id2.5}
    \ee
Hence,
    \be
    \rh_2=\frac{i}{4}\,[\nu(\rx),Q_1]=\frac{i}{4}[\eta_{+1},\nu(\rx)],
    \label{id3}
    \ee
where we have also made use of the first equation in (\ref{Q12}).
We should also mention that the identities~(\ref{id1}) and
(\ref{id2.5}) can be directly obtained from the pseudo-Hermiticity
condition (\ref{ph}) by substituting (\ref{exp}) in (\ref{ph}) and
using (\ref{cbh}) and (\ref{expand}).

We can easily use (\ref{eta-1-xy}) and (\ref{id3}) to yield the
expression for the integral kernel of $\rh_2$, namely
    \be
    \br\rx|\rh_2|\ry\kt={\mbox{\small$\frac{1}{32}$}}\,
    (4+2|\rx+\ry|-|\rx+\ry+2|-|\rx+\ry-2|)\,
    {\rm sign}(\rx-\ry)[\nu(\rx)-\nu(\ry)],
    ~~~~~~~~\forall \rx,\ry\in\R.
    \label{h2-xy}
    \ee
As seen from this equation,
    \be
    \br\rx|\rh_2|\ry\kt=0,~~~~{\rm if}~~~~
    \rx\notin [-1,1]~~{\rm and}~~\ry\notin [-1,1].
    \label{h2=zero}
    \ee

We can express $\rh_2$ as a function of the $\rx$ and $\rp$ by
performing a Fourier transformation on the $\ry$ variable
appearing in (\ref{h2-xy}), i.e., computing
    \be
    \br\rx|\rh_2|\rp\kt:=(2\pi)^{-1/2}\int_{-\infty}^\infty
    \br\rx|\rh_2|\ry\kt \, e^{i\rp \ry}d\ry.
    \label{F-trans}
    \ee
This yields $\rh_2$ as a function of $\rx$ and $\rp$, if we order
the factors by placing $\rx$'s to the left of $\rp$'s. We can
easily do this by expanding $\br\rx|\rh_2|\rp\kt$ in powers of
$\rp$. Denoting the $\rx$-dependent coefficients by $\omega_n$, we
then have
    \be
    \rh_2=\sum_{n=0}^\infty \omega_n(\rx)\,\rp^n,
    \label{expand-h2}
    \ee
where we have made the implicit assumption that
$\br\rx|\rh_2|\rp\kt$ is a real-analytic function of $\rp$.

The Fourier transform of $\br\rx|\rh_2|\ry\kt$ can be performed
explicitly.\footnote{One way of doing this is to use the integral
representations of the absolute value and sign function, as given
in (\ref{identities}), to perform the $\ry$-integrations in
(\ref{F-trans}) and use the identities
    \[ \int_{-\infty}^\infty \frac{e^{ia
    u}du}{u(u-k)}=\frac{i\pi}{k}(e^{iak}-1)\,{\rm sign}(a),~~~~
    \int_{-\infty}^\infty \frac{e^{ia
    u}du}{u(u-k)^2}=\frac{i\pi}{k^2}[1+(iak-1)e^{iak}]\,{\rm sign}
    (a),~~~~~~~\forall a,k\in\R,\]
to evaluate the remaining two integrals. The resulting expression
is too lengthy and complicated to be presented here.} We have
instead used Mathematica to calculate $\br\rx|\rh_2|\rp\kt$ and
found the coefficients $\omega_n$ for $n\leq 5$. It turns out that
indeed $\br\rx|\rh_2|\rp\kt$ does not have a singularity at
$\rp=0$, and that $\omega_0, \omega_2,\omega_4$ are real and
vanish outside $(-3,3)$ while $\omega_1,\omega_3,\omega_5$ are
imaginary and proportional to $\theta(\rx)-1/2$ outside $(-3,3)$.
As we will explain momentarily these properties are necessary to
ensure the Hermiticity of $\rh$.

Figures~1, 2 and 3 show the plots of real part of $\omega_n$ for
$n=0,2,4$ and the imaginary part of $\omega_n$ for $n=1,3,5$.
    \begin{figure}[p]
    \centerline{\epsffile{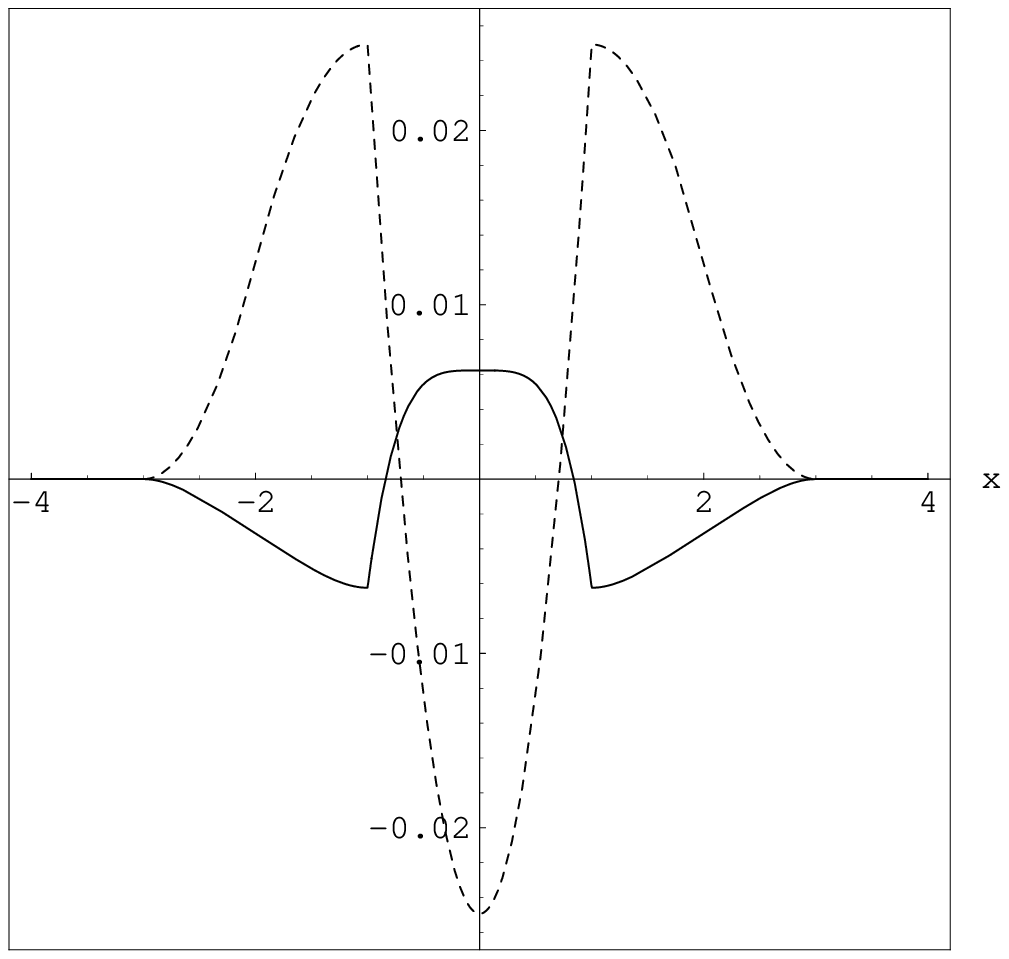}}
    \centerline{
    \parbox{14cm}{\caption{Graph of the real part of $\omega_0$
    (dashed curve) and $\omega_2$ (full curve).}\label{fig1}}}
    \vspace{1cm}
    \centerline{\epsffile{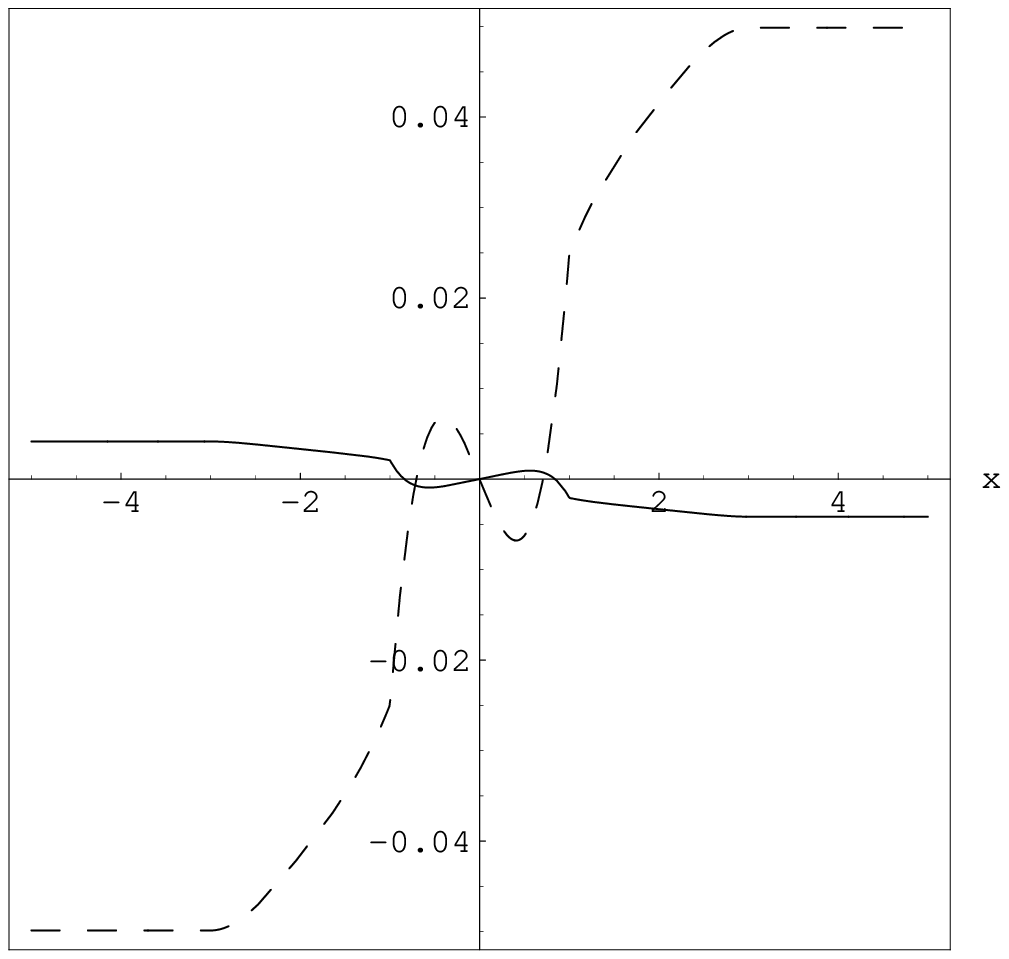}}
    \centerline{
    \parbox{14cm}{\caption{Graph of the imaginary part of $\omega_1$
    (dashed curve) and $\omega_3$ (full curve).}\label{fig2}}}
    \end{figure}
    \begin{figure}
    \centerline{\epsffile{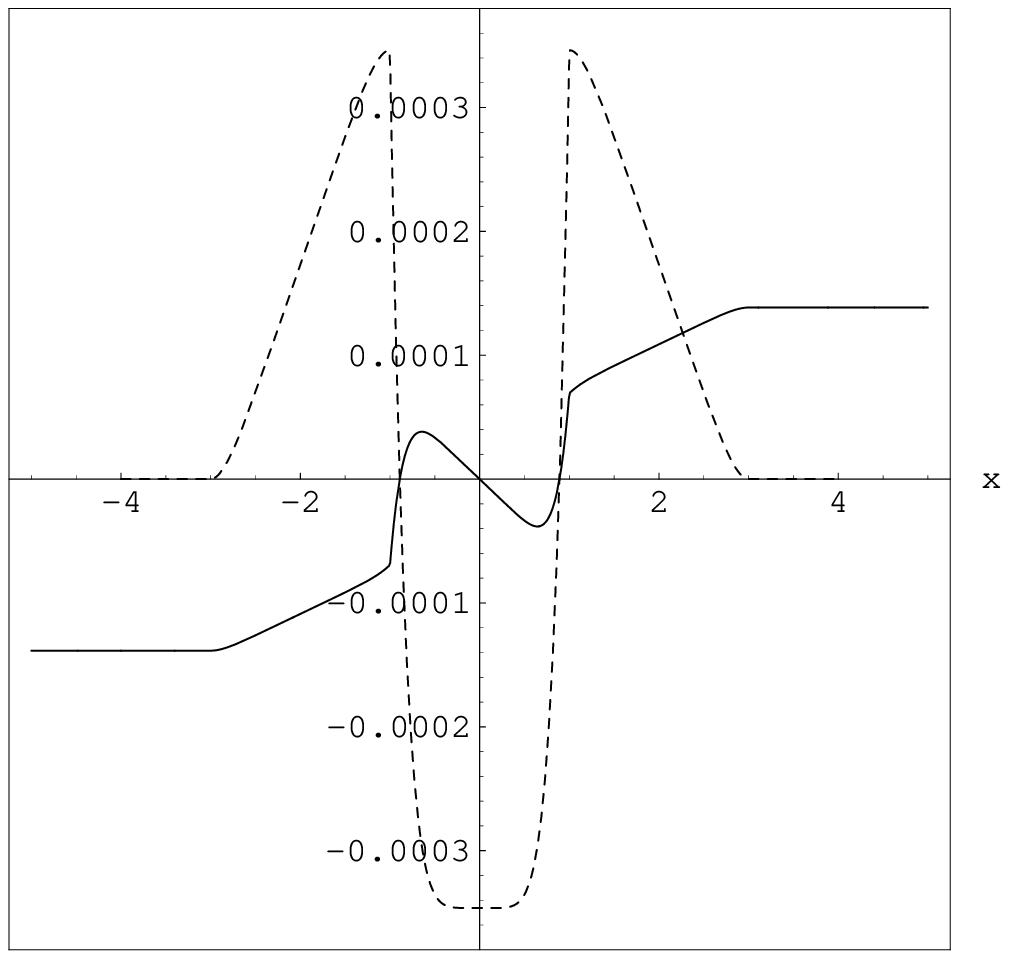}}
    \centerline{
    \parbox{14cm}{\caption{Graph of the real part of $\omega_4$
    (dashed curve) and the imaginary part of $\omega_5$
    (full curve).}\label{fig3}}}
    \end{figure}
As seen from these figures (the absolute value of) $\omega_n$
sharply decreases with $n$, which suggests that a truncation of
(\ref{expand-h2}) yields a a good approximation for the action of
$\rh_2$ on the wave functions with bounded and sufficiently small
$\rx$-derivatives.

If we use $\br\rp|\rh_2|\rx\kt=\br\rx|\rh_2|\rp\kt^*$ to determine
the form of $\rh_2$ and suppose that $\omega_{2n}(\rx)$ are real
and $\omega_{2n+1}(\rx)$ are imaginary for all $n=0,1,2,3,\cdots$,
we find
    \[\rh_2=\sum_{n=0}^\infty \rp^n\omega_n(\rx)^*=
            \sum_{n=0}^\infty [\rp^{2n}\omega_{2n}(\rx)-
            \rp^{2n+1}\omega_{2n+1}(\rx)].\]
Adding both sides of this relation to those of (\ref{expand-h2})
and diving by two, we obtain
    \be
    \rh_2=\frac{1}{2}\,\sum_{n=0}^\infty
    \{a_n(\rx),\rp^{2n}\},~~~~~~~~~
    a_n(\rx):=\omega_{2n}(\rx)+i\omega'_{2n+1}(\rx),
    \label{confine}
    \ee
where $\{\cdot,\cdot\}$ stands for the anticommutator, a prime
denotes a derivative, and we have made use of the identity:
$[f(\rx),\rp^m]=\{if'(\rx),\rp^{m-1}\}$. It is important to note
that because $\omega_{2n}(\rx)$ are real and $\omega_{2n+1}(\rx)$
are imaginary, $a_n(\rx)$ are real. Moreover, outside $(-3,3)$,
$\omega_{2n}(\rx)$, $\omega_{2n+1}'(\rx)$, and consequently $a_n$
vanish. Therefore, we can express $\rh_2$ in the manifestly
Hermitian form~(\ref{confine}) with all the $\rx$-dependent
coefficient functions vanishing outside $(-3,3)$. Figure~4 shows
the plots of $a_n$ for $n=0,1,2$. They are all even functions of
$\rx$ with an amplitude of variations that decreases rapidly as
$n$ increases.
    \begin{figure}[p]
    \centerline{\epsffile{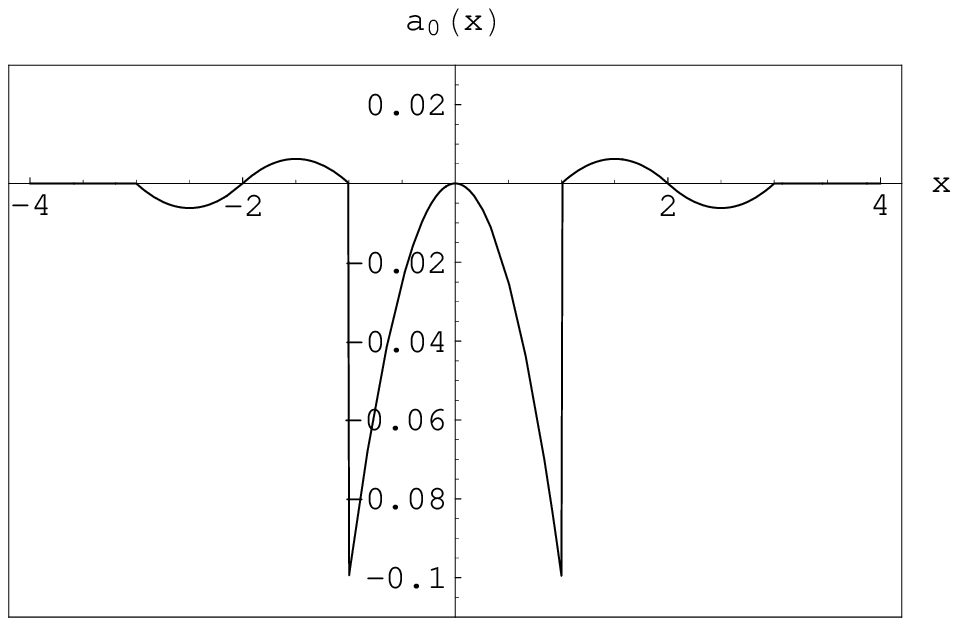}}
    \vspace{1.5cm}
    \centerline{\epsffile{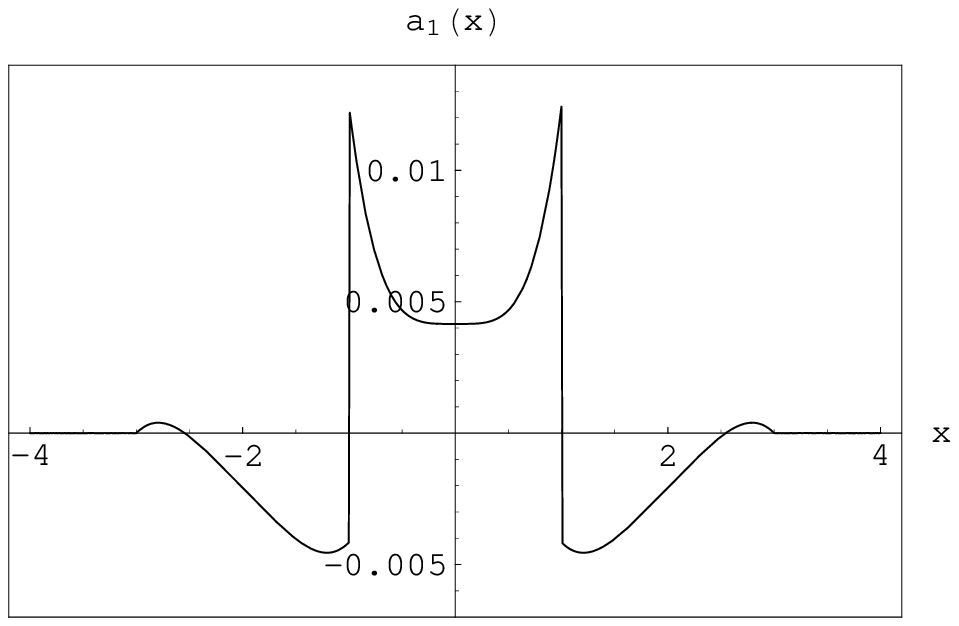}}
    \vspace{1,5cm}
    \centerline{\epsffile{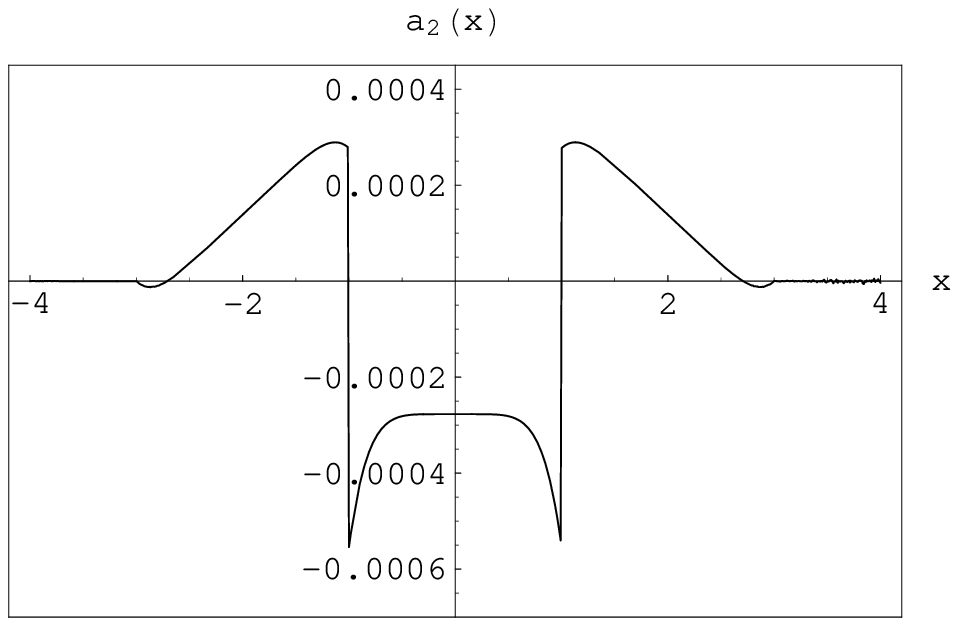}}
    \centerline{
    \parbox{14cm}{\caption{Graph of $a_0$, $a_1$ and $a_2$}
    \label{fig4}}}
    \end{figure}

Next, we scale back the relevant quantities and use
(\ref{scale1}), (\ref{rh}), (\ref{h=h2}), and (\ref{confine}) to
obtain
    \be
    h=\frac{p^2}{2m}+\frac{\zeta^2}{2}\sum_{n=0}^\infty
    \{\alpha_n(x),p^{2n}\}+{\cal
    O}(\zeta^3),~~~~~~~~~~~~~~~
    \alpha_n(x):=2m \left(\mbox{$\frac{L}{2\hbar}$}\right)^{2(n+1)}
    a_n(\mbox{\small $\frac{2x}{L}$ }).
    \label{unscaled-h}
    \ee
In view of the fact that $a_n$ and $\alpha_n$ are real-valued even
functions, $h$ is a manifestly Hermitian ${\cal P}$- and ${\cal
T}$-symmetric Hamiltonian. We can also express it in the form
    \be
    h=\frac{1}{4}\{m^{-1}_{\rm eff}(x),p^2\}+ w(x)+
    \frac{\zeta^2}{2}\sum_{n=2}^\infty
    \{\alpha_n(x),p^{2n}\}+{\cal
    O}(\zeta^3),
    \label{eff}
    \ee
where
    \[ m_{\rm eff}(x):=\frac{m}{1+2m\zeta^2\alpha_1(x)},~~~~~~~~~~~~~
    w(x):=\zeta^2\alpha_0(x).\]
Therefore, for low energy particles where one may neglect terms
involving 4th and higher powers of $p$, the Hamiltonian $h$ and
consequently $H$ describe motion of a particle with an effective
position dependent mass $m_{\rm eff}(x)$ that interacts with the
potential $w(x)$. Figure~5 shows a graph of $m_{\rm eff}(x)$ for
$m=1/2,\hbar=1,L=2$ and $\zeta=1/3$. For the same values of these
parameters, $w(x)=a_0(x)/9$. See Figure~4 for a graph of $a_0$.
\begin{figure}[ht]
    \centerline{\epsffile{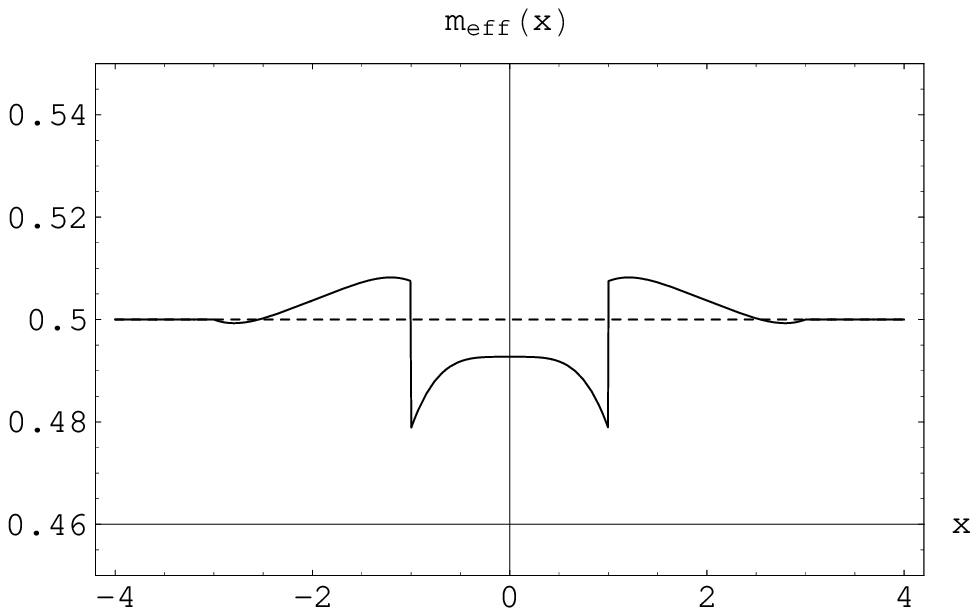}}
    \centerline{
    \parbox{12cm}{\caption{Graph of the effective mass $m_{\rm eff}$
    (full curve) for $m=\mbox{\small$\frac{1}{2}$},\hbar=1,L=2$ and
    $\zeta=\mbox{\small$\frac{1}{3}$}$. The dashed curve represents
    $m=\mbox{\small$\frac{1}{2}$}$}\label{fig5}}}
    \end{figure}

If we replace $(x,p)$ of (\ref{unscaled-h}) and (\ref{eff}) with
their classical counterparts $(x_c,p_c)$, we obtain the
`classical' Hamiltonian:
    \be
    \tilde H_c=\frac{p_c^2}{2m}+\frac{\zeta^2}{2}\sum_{n=0}^\infty
    \alpha_n(x_c)\:p_c^{2n}+{\cal O}(\zeta^3)=
    \frac{p_c^2}{2m_{\rm eff}(x_c)}+w(x_c)+
    \frac{\zeta^2}{2}\sum_{n=2}^\infty
    \alpha_n(x_c)\:p_c^{2n}+{\cal O}(\zeta^3),
    \label{class-H}
    \ee
which coincides with the free particle Hamiltonian outside the
{\em physical interaction region}, i.e., $(\mbox{\small
$-\frac{3L}{2}$},\mbox{\small $\frac{3L}{2}$})$. The fact that
this region is three times larger than the support $(\mbox{\small
$-\frac{L}{2}$},\mbox{\small $\frac{L}{2}$})$ of the potential
$v(x)$ is quite surprising. Note also that $\tilde H_c$ is an even
function of both the position $x_c$ and momentum $p_c$ variables.

Figure~6 shows the phase space trajectories associated with the
Hamiltonian $\tilde H_c$ for $L=2$, $\hbar=1$, $m=1/2$,
$\zeta=Z=1/3$.
    \begin{figure}[ht]
    \centerline{\epsffile{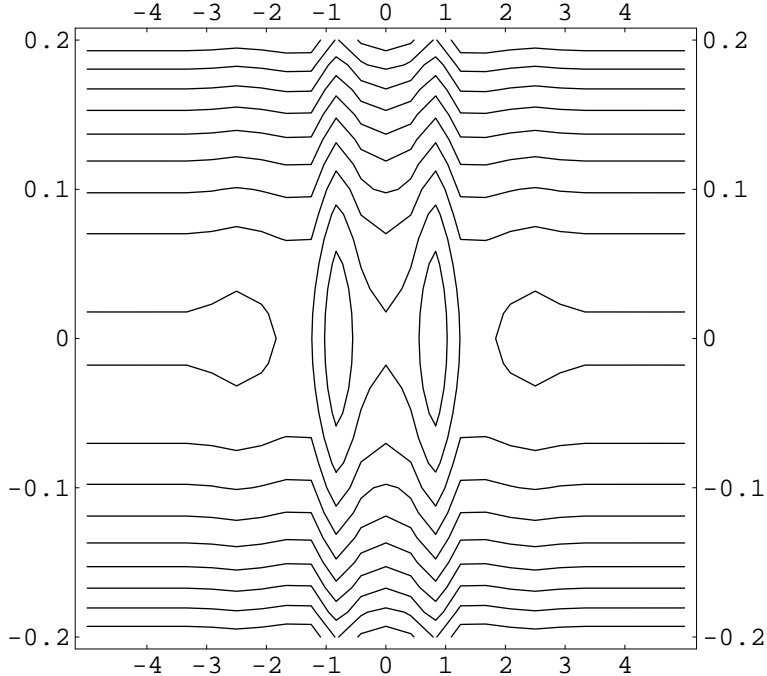}}
    \centerline{
    \parbox{12cm}{\caption{Phase space trajectories of the
    Hamiltonian $\tilde H_c(x_c,p_c)$ for $m=\mbox{\small$\frac{1}{2}$},\hbar=1,L=2$ and
    $\zeta=\mbox{\small$\frac{1}{3}$}$. The horizontal and vertical
    axes are respectively those of $x_c$ and $p_c$.}\label{fig6}}}
    \end{figure}
For large values of the momentum the trajectories are open curves
describing the scattering of a particle due to an interaction that
takes place within the physical interaction region, $(-3,3)$. For
sufficiently small values of the momentum closed trajectories are
generated. These describe a particle that is trapped inside the
physical interaction region. This is consistent with the fact that
for small $p_c$, $\tilde H_c$ is dominated by the potential term
$w(x_c)$ which in view of its relation to $a_0(x)$ and Figure~4
can trap the particle.

We wish to emphasize that because we have not yet takes the
$\hbar\to 0$ limit of $\tilde H_c$, we cannot identify it with the
the true classical Hamiltonian $H_c$ for the quantum Hamiltonian
$h$ and consequently $H$. Given the limitations of our
perturbative calculation of $\tilde H_c$, we are unable to
determine this limit.\footnote{This is in contrast with both the
${\cal PT}$-symmetric square well and the ${\cal PT}$-symmetric
cubic anharmonic oscillator studied in \cite{jpa-2004c}  and
\cite{p64}, respectively. In the former system the presence of an
exceptional spectral point imposes the condition that $\zeta$ must
be of order $\hbar^2$ or higher and consequently the classical
system is the same as that of the Hermitian infinite square well
\cite{jpa-2004c}. In the latter system, the $\hbar\to 0$ limit of
the associated Hermitian Hamiltonian can be easily evaluated and
classical Hamiltonian obtained \cite{p64}.} Therefore, we cannot
view the presence of closed phase space trajectories for $\tilde
H_c$ as an evidence for the existence of bound states of $h$ and
$H$. This is especially because these trajectories are associated
with very low momentum values where the quantum effects are
expected to be dominant.

\section{Conclusion}

In this paper we explored for the first time the utility of the
methods of pseudo-Hermitian quantum mechanics in dealing with a
non-Hermitian ${\cal PT}$-symmetric potential $v(x)$ that has a
continuous spectrum. Using these methods we were able to obtain
the explicit form of the metric operator, the pseudo-Hermitian
position and momentum operators, the localized states, and the
equivalent Hermitian Hamiltonian perturbatively.

Our analysis revealed the surprising fact that the physical
interaction region for this model is three times larger than the
support of the potential, i.e., there is a region of the
configuration space in which $v(x)$ vanishes but the interaction
does not seize.

A simple interpretation for this peculiar property is that the
argument $x$ of the potential $v(x)$ is not a physical observable
and the support $(\mbox{\small $-\frac{L}{2}$},\mbox{\small
$\frac{L}{2}$})$ of $v(x)$ being a range of eigenvalues of $x$
does not have a direct physical meaning. This observation
underlines the importance of the Hermitian representation of
non-Hermitian (inparticular ${\cal PT}$-symmetric) Hamiltonians
having a real spectrum.

The Hermitian representation involves a nonlocal Hamiltonian that
is not suitable for the computation of the energy spectrum or the
$S$-matrix of the theory. Yet it provides invaluable insight in
the physical meaning and potential applications of
pseudo-Hermitian and ${\cal PT}$-symmetric Hamiltonians and is
indispensable for the determination of the other observables of
the corresponding quantum systems.

{

}

\ed